\documentclass[aps,prl,amsmath,amssymb,twocolumn,groupedaddress,notitlepage,showpacs,floatfix,superscriptaddress]{revtex4-1}

\usepackage{graphicx}
\usepackage{bm,color}

\usepackage{amsmath}
\usepackage{bm,color}
\usepackage{multirow}
\usepackage{ulem}
\usepackage{xcolor}

\begin{document}

\title{Plateau transitions of spin pump and bulk-edge correspondence}
\date{\today}

\author{Yoshihito Kuno}
\author{Yasuhiro Hatsugai}
\affiliation{Department of Physics, University of Tsukuba, Tsukuba, Ibaraki 305-8571, Japan}

\begin{abstract}
Sequential plateau transitions of quantum spin chains ($S$=1,3/2,2 and 3) are demonstrated by a spin pump using dimerization and staggered magnetic field as synthetic dimensions. 
The bulk is characterized by the Chern number associated with the boundary twist and the pump protocol as a time. It counts the number of critical points in the loop that is specified by the $Z_2$ Berry phases.
With open boundary condition, discontinuity of the spin weighted center of mass due to emergent effective edge spins also characterizes the pump as the bulk-edge correspondence. 
It requires extra level crossings in the pump as a super-selection rule that is consistent with the Valence Bond Solid (VBS) picture.
\end{abstract}

\maketitle
\section{Introduction}
Integer $S$ spin systems have been extensively studied after Haldane proposed an exotic proposal for the uniform Heisenberg chains, the ground state is gapped if the spin is integer \cite{Haldane}.
Various numerical studies support this conjecture positively \cite{Kennedy, Hatsugai1992,White,Nakano2009,Nakano2018}.
This Haldane conjecture is proved for
the VBS states
of the AKLT model \cite{Affleck1987,Affleck1988,Affleck1989}. With open boundaries for $S=1$ case, there appears extra low energy level structure within the Haldane gap 
which is described by emergent effective $S=1/2$ spins at both ends of the chain \cite{Kennedy}. 
This is also consistent with the VBS picture \cite{Affleck1987,Affleck1988,Affleck1989}. 
The edge states were also experimentally observed \cite{Hagiwara}. 
According to the bulk-edge correspondence, these effective degrees of freedom is due to the non-trivial bulk \cite{Hatsugai1993}.

This Haldane spin chain is an example of symmetry protected topological (SPT) phases \cite{Pollmann2010,Chen,Pollmann2012}. 
It is robust against disorder or deformations as far as the symmetry is preserved. 
The SPT phase of the integer $S=1,2$ dimerized spin chain has been characterized by $Z_2$ Berry phases and is consistently understood by the VBS picture \cite{Hirano2008_2,Hirano2008,Katsura2007,Mila,Fubasami}.

Recently, topological charge pump (TCP) \cite{Thouless} is a hot topic again in the condensed matter community since recent artificial quantum systems realized it experimentally \cite{Lohse,Nakajima,Schweizer,Kraus_ex,Ozawa,Cooper}. 
The bulk-edge correspondence of the TCP has been also studied recently \cite{Hatsugai2016,KH2020}, although its bulk description is old. 
{A recent field theoretical study has also discussed the generalized berry phases related to the TCP \cite{Po-Shen2020}.} 
The TCP of fermionic/bosonic systems \cite{Wang,RLi,YKe,Kuno2017,Nakagawa,Hayward,Greschner} and spin pump for $S=1/2$ have been discussed \cite{Hu,Shindou}, which is characterized by the Chern number \cite{TKNN}. 
Still, nature of the spin pump for $S\ge 1$, especially bulk-edge correspondence, remains unclear.

In this letter, we clarify the presence of a nontrivial topological spin pump in a dimerized Heisenberg model with generic $S$ ($S=1,3/2,2$ and $3$). 
Without using a dimensional reduction from the historical ancestor, that is, the quantum Hall system in 2D \cite{Kraus}, we are proposing a non-trivial topological pump connecting two SPT phases with different edge states. 
We have demonstrated this spin pump by calculating Chern number of the bulk and also discussed low energy spectrum of the SPT phases with small symmetry breaking parameters, that determines the behavior of the edge states during the pump. 
This is reflected by a series of discontinuities of the spin weighted center of mass (sCoM). 
Using the numerical data, 
we have demonstrated the bulk-edge correspondence of the generic spin chains and its relation to the VBS picture have been clarified. 
It suggests inevitable existence of the low energy boundary degrees of freedom in the infinite system.
\begin{figure*}[t]
\centering
\includegraphics[width=18cm]{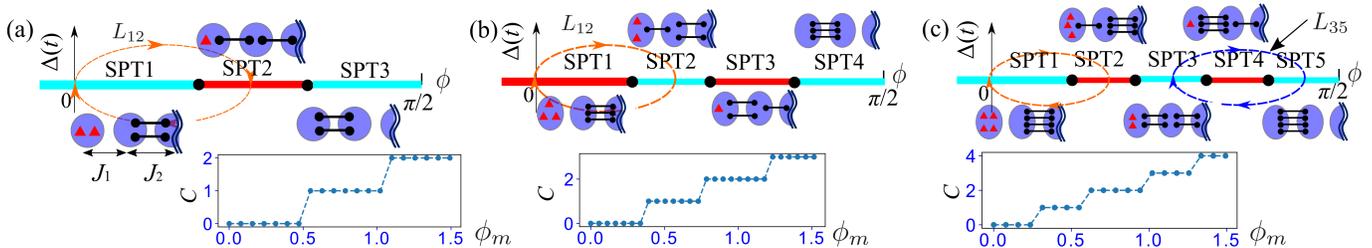}
\caption{(a) Schematic figures for $S=1$ SPT phase diagram, VBS pictures with left open boundary and a pump protocol in $\phi-\Delta$ plane.  
The SPT phase colored in light blue (red) has $0$($\pi$)-Berry phase for $J_2$-link. Here, $\phi=0$ is {\it not} a transition point. In the VBS picture, the red triangle represents spin $1/2$. The bottom panel is the plateau transition. 
(b) Schematic figure for $S=3/2$ SPT phase diagram, the VBS pictures and a pump protocol in $\phi-\Delta$ plane. The bottom panel is plateau transition.
(c) Schematic figure for $S=2$ SPT phase diagram, the VBS pictures. The blue loop represents the pump protocol $L_{35}=L_{15}-L_{13}$, which gives $C_{35}=C_{15}-C_{13}=2$.
The bottom panel is the plateau transition.}
\label{Fig1}
\end{figure*}

\section{Model}
In this Letter, we consider a dimerized Heisenberg model with generic $S$ \cite{Nakamura,Hirano2008}, 
$H_{DH}=\sum^{L-1}_{j=0}J_{j}\vec{S}_{j}\cdot\vec{S}_{j+1},
\:\:\:S_j^2 =S(S+1)$,
where $\vec{S}_{j}=(S^{x}_{j}, S^{y}_{j},S^{z}_{j})$ and $J_j$ is dimerized, $J_{j\in even}=J_1$ and $J_{j\in odd}=J_2$ \cite{exp_model}. 
The phase diagrams for $S=1/2$ and $3/2$ \cite{Yajima,Yamamoto,Kitazawa} and
integer cases ($S=1$ and $2$) \cite{Hirano2008,Katsura2007} have been discussed before. There are sequence of gapped SPT phases denoted by SPT1, SPT2, $\cdots$, that appear by changing the ratio $J_1/J_2$. The schematic phase diagrams for $S=1,3/2,2$ are shown in Fig.~\ref{Fig1}. 
All of the SPT phases of the $H_{DH}$ are
protected by one of $D_{2}$, time-reversal and bond-centered inversion symmetries \cite{Pollmann2010,Pollmann2012,remark_symmetry}. 
The state of the SPT phases in $H_{DH}$ with generic $S$ is discussed by $Z_2$ Berry phases protected by the symmetry \cite{Hirano2008}.

The SPT phases of the bulk are characterized by the Berry phase by a local twist \cite{EPL-YHIM,Hatsugai2005,Hatsugai2006,Hatsugai2007,Hatsugai2011,PRL-TK-TM-YH}, $\frac{J_2}{2}(e^{i \theta }S^{+}_0S^{-}_{L-1}+e^{- i \theta }S^{-}_0S^{+}_{L-1})$, ($e^{i \theta }\in S^1$, $\theta\in (\pi,\pi]$). 
Here we assume the sites are  labeled by $i=0,\cdots,L-1$.
The Berry phase is given by 
$i\gamma=\int_{S^1} A_{\theta}(\theta)d\theta $ where $A_{\theta}(\theta) =
\langle G(\theta )|\partial_{\theta}G(\theta)\rangle$ 
and $|G(\theta)\rangle$ is the ground state of $H_{DH}(\theta)$. 
The SPT phases for $S=1,2$ are discussed and
characterized by the quantized value of
$\gamma=0,\pi\in\mathbb{Z}_2$ that is consistently
understood by the VBS picture
by assigning $\pi$ for the valence bond (VB) \cite{Hirano2008,Katsura2007}.
Note that the gap of a finite system
remains open for a system with periodic boundary condition
even near the transition points. With twisted boundary condition, however,
the gap vanishes mostly at $\theta =\pi$. It is due to the topological charge
of the two bulks characterized by different $Z_2$ Berry phases.
It suggests that extending a system by adding $U(1)\in S^1$ twist
as an associate dimension 
is useful for the topological transition for finite systems.
This $S^1$ is small in the sense that the effects are infinitesimal 
in infinite systems \cite{NTW,Kudo}.
In this sense, transition points of the $S^1$-enlarged system
is the same as that of the original one. 
Symmetry sometimes requires gap closing for the $S^1$-enlarged system.
The gap necessarily closes for a translationally invariant half-integer spin
chain as an analogue of the Lieb-Schultz-Mattis theorem for
the extended system \cite{Hirano2008_2}.

For each SPT phases in Fig.~\ref{Fig1} (a), (c) and (e), the value of $\gamma$ is consistently understood by the VBS picture. 
Since the Berry phase has modulo $2\pi$ ambiguity, the number of bonds is specified in modulo 2.
The VBS picture actually works more than that as shown later. 
In the topological spin pump we propose, we have observed emergent edge states predicted by the VBS picture. 
According to the bulk-edge correspondence, this is specified by the Chern number in the extended parameter space (See below). 
In the phase diagrams in Fig.~\ref{Fig1} (a), (c) and (e), all critical transition points on $\Delta=0$ are gapless for an infinite system. 
Even in a finite system, the gap closes as an $S^1$-enlarged system.

\section{Topological plateau transition of pump for
generic spin} 
To characterize the ground state of $H_{DH}$, 
let us introduce a symmetry breaking term, $H_{SB}=\Delta(t)\sum_{j}(-1)^{j+1}S^{z}_j$,
and consider an extended Hamiltonian $H=H_{DH}+H_{SB}$
where $\Delta(t)$ is a periodic dynamical parameter with a period $T$, 
which breaks {\it all} symmetries protecting the SPT phases. The Hamiltonian $H$ preserves $U(1)$ symmetry of the subgroup of $SU(2)$, which implies conservation of total $S^{z}$. 
Here, a pump protocol is specified by a periodic modulation of the parameters. 
To be specific, we take $J_1(t)=\sin\phi(t)$, $J_2(t)=\cos\phi(t)$ with $\phi(t)=\phi_{m}[1-\cos(2\pi t/T)]/2$ and $\Delta(t)=\sin (2\pi t/T)$. 
The amplitude of the modulation, $\phi_m$, is chosen so that the ground state of $H(t)$ at $t=0,T/2$ (where $H_{SB}=0$) belong to the different SPT phases as 
$(J_1(t),J_2(t))=(0,1),(\sin \phi_m,\cos \phi_m)$, $\Delta =0$ for $t=0$ and $T/2$, respectively. 
Note that the gap always remains open and the ground state is unique in the pump as for a periodic system \cite{KH2020}.
  
The pump protocol is characterized by the (spin) Chern number \cite{Arovas} of the periodic system with local boundary twist $e^{i\theta}$, {$C=\frac{1}{2\pi i}\int^{T}_{0}dt\int^{2\pi}_{0}d\theta B$,} where {$B= \partial _\theta A_t -\partial _t A_\theta$}, {$A_\alpha = \langle g|\partial _\alpha g \rangle $, $\alpha =\theta ,t$ 
where $|g\rangle $}
is a gapped and unique ground state of $H$. 
The Berry phase defined at $t=0$ and $T/2$, $|G \rangle =| g \rangle \big|_{\Delta =0}$, is quantized \cite{Shindou}.

This Chern number coincides to the total pumped spin of the bulk \cite{Hatsugai2016,KH2020} (See also Sec.I in supplemental material \cite{Sup}). 
Here let us define a Chern number, $C_{1k}$, for the protocol specified by the loop $L_{1k}$ starting from the SPT1 and passing through the other $k$-th SPT (SPT$k$) phases as shown in Fig.~\ref{Fig1} (a), (c) and (e). We have used the formula \cite{FHS2005} by diagonalizing the system \cite{Quspin} with even number of spins within the total $S^z\equiv \sum^{L-1}_{j=0}S^z_j=0$ sector since the ground state is unique. The Berry phase at $t=0$ is $2\pi S$ since the dimers are decoupled and the twist is gauged out \cite{Hirano2008}. 
Let us define a path $L_{kk ^\prime}$ starting from the SPT$k$ and passing through the SPT$k ^\prime$ and the corresponding Chern number $C_{kk ^\prime }$ (See the blue loop in Fig.~\ref{Fig1} (e) as an example). 
Since the path can be deformed into the two paths $L_{1k}$ and $L_{1 k ^\prime }$ as $L_{k k ^\prime }=L_{1k ^\prime }-L_{1 k}$ without gap closing, the Chern numbers satisfy the relation, $C_{kk ^\prime } = C_{k ^\prime } -C_{k}$. 

The results of $C_{1k}$ for $S=1$, $3/2$ and $2$ cases are plotted in Fig.~\ref{Fig1} (b), (d) and (f) \cite{remark0}.
By changing the width of modulation of the dimerization $\phi_m$, the Chern number changes step by step, it is an analogue of the quantum Hall plateau transitions \cite{Kivelson,Hatsugai1999,Kagalovsky,Kawarabayashi,Morita}. 
The maximum Chern number for each case is $C=2S$, that corresponds to the total number of possible dimerization transitions.
The case for $S=3$ is similar (see Sec.II in \cite{Sup}). 
The Chern number of the bulk for the pump protocol is specified by the topology of the two SPT phases where the pump are passing through. 
This is clear considering a system with edges as discussed later. 
The gap closing points of the SPT phases on the $\Delta =0$ line are topological obstruction for the loop specified by the pump protocol on the $\phi-\Delta $ plane (strictly speaking, the obstruction is for the $S^1$-enlarged system.) 
Therefore only when the loop passes through these points, the Chern number is allowed to change.
The Chern number of a generic pump is given by a sum of the Chern numbers of the critical points inside the protocol loop \cite{remark10}. 
The plateau transition of the pump in 2D is induced by the SPT transition of the 1D spin chain \cite{remark11}. 

\section{Spin center of mass with open boundary}
Let us investigate the topological spin pump with open boundary, especially, properties of edge states. 
To this end, we employ the density matrix renormalization group method in TeNPy package \cite{TeNPy}. 
We consider the sCoM \cite{Hatsugai2016} given by 
$P(t)=\sum^{L-1}_{j=0}\langle g(t)|x_jS^{z}_{j}|g(t)\rangle$, 
where $j_0=(L-1)/2$, $x_j=(j-j_0)/{L}\in(-1/2,1/2)$. 
This sCoM gives a spin current $J=\partial _t P$. 
Note that the sCoM is only well defined for a system with boundaries 
and is not well defined for a system with periodic boundary condition. 
Since the pump is periodic in time, 
the sCoM, $P(t)$, is also. 
It implies 
$
0 = \int_{0}^T dt\, \partial_t P
=\sum_i\int_{t_{i-1}}^{t_i}dt\, \partial_t P +P(t)|^{t_i+0}_{t_i-0}
$
where $P$ is piecewise continuous and have  discontinuities at $t=t_i$, ($i=1,2,\cdots$) (periodicity in time is assumed for the summation) \cite{Hatsugai2016}. 
Then, for any path passing through SPT$k$ and SPT$k ^\prime$ in the parameter space, the pumped spin $Q^e_{kk ^\prime}$ in the cycle 
by bulk for a system with open boundary condition is related to the sum of the discontinuities
\begin{align}
Q^e_{k k ^\prime } =\sum_i\int_{t_{i-1}}^{t_i}dt\, \partial_t P= I_{k k ^\prime },
\label{eq:bulk-charge}
\end{align}
where $I_{k k ^\prime }\equiv-\sum_iP(t)\big|^{t_i+0}_{t_i-0}$.

In the spin model in this Letter, each discontinuity is $\pm 1$ which is induced by exchanging the left and right edge states \cite{Hatsugai2016}.
It is due to the symmetry of the system. 
In a generic situation, single annihilation (creation) of the edge state is allowed as $P(t)\big|^{t_i+0}_{t_i-0}=+1/2$ at the left (right) boundary and $-1/2$ at the right (left).

\begin{figure}[t]
\centering
\includegraphics[width=8.5cm]{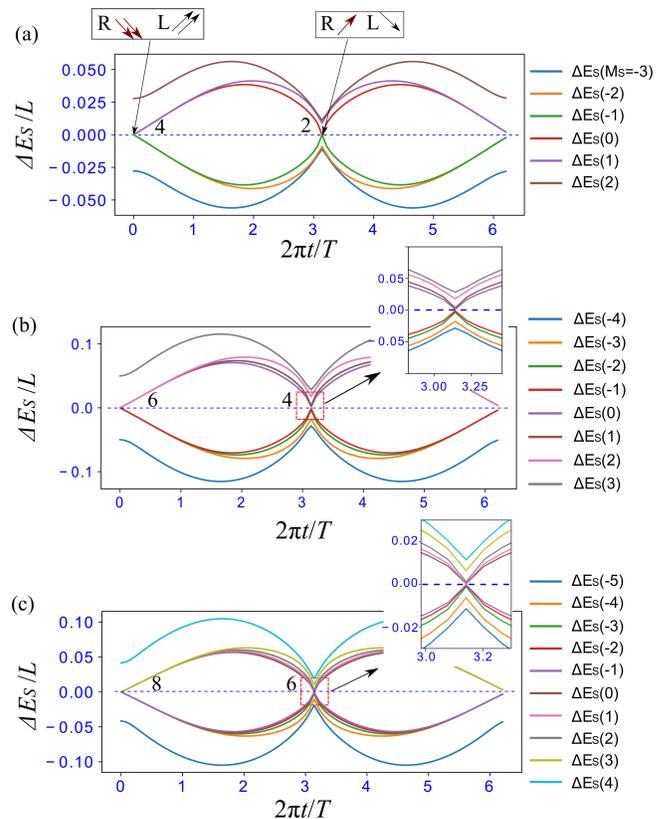}
\caption{(a) $\Delta E_S$ for $S=1$ case. We set $L=36$. 
The label `R' and `L' represent right and left edge states. 
The upward and downward arrows represent the direction when the edge states cross the blue dashed line.
(b) $\Delta E_S$ for $S=3/2$ case. We set $L=32$. 
(c) $\Delta E_S$ for $S=2$ case. We set $L=24$. 
The numbers at level crossings indicate the number of the degeneracy of $\Delta E_S$.}
\label{Fig2}
\end{figure}

We also calculate an excitation energy \cite{Hu} defined by
$\Delta E_S(M_S,t)=E_S(M_S+1,t)-E_S(M_S,t)$ for each $M_S$ sector
where $E_S(M_S,t)$ is the ground state energy of $H$ within a subspace total $S^{z}=M_S$.
Let us discuss $\Delta E_S(M_S,t)$ at $M_S=0,\pm 1,\cdots$. 
We choose the same parameters as shown in the bulk calculation of Fig.~\ref{Fig1} 
where the pumped spin $C_{1k}=1$. The amplitude of the modulation, $\phi_m$,
is set to connect the SPT1 to the midpoint of the SPT2 phase 
($\phi_m=\pi/4,3\pi/16,\pi/8$ for $S=1,3/2,3$ respectively). 
The results for $S=1$, $3/2$ and $2$ are shown in Fig.~\ref{Fig2} (a), (b) and (c). 
Numerically calculated excitation gaps become very small at $t=0$ and $t=T/2$. 
These extremely small gaps are due to the interaction between the edge states localized near both ends of the system as emergent degrees of freedom associated with non-trivial bulk \cite{remark5}. 
It is an extension of the well known effective $S=1/2$ boundary degrees of $S=1$ case at $t=T/2$, 
that makes four-fold degeneracy in the infinite system \cite{Kennedy}. 
The symmetry breaking term $H_{SB}$ makes the degeneracy to the level crossings observed in Fig.~\ref{Fig2}. 
The degeneracy at $t=0$ are trivial and is given by the addition of the bare spin $S$ at the boundaries ($S\otimes S=2S\oplus\cdots \oplus 0$).
They are $(2S+1)^2$-fold in total and $\Delta E_S(M_S,0)=0$ for $4S$ different $M_S$ sectors $M_S=-2 S,\cdots,2 S-1$. Then the discontinuity at $t=0$ is
$-\sum_iP(t)\big|^{+0}_{-0}=2S$.

Although the degeneracy at $t=T/2$ is non-trivial, 
it is consistently explained by the VBS picture \cite{Hirano2008,Katsura2007,Hirano2008_2}. 
See Fig.~\ref{Fig1} (a), (c) and (e). 
Based on the VBS picture, in the SPT2 phase, a single valence bond (VB) connected to the neighboring spin, 
reduces the spin by $1/2$ at the boundary spins. 
It implies the effective free spins at the boundaries are $S_{{\rm eff}}=S-1/2$. 
For $S=1$ case at $t=T/2$, it implies total degeneracy of $(2S_{{\rm eff}}+1)^2=4$ and $\Delta E_S(M_S,T/2)=0$ for $4S_{{\rm eff}}=2$ different $M_S$ sectors. 
The same can be true for the SPT$k$ phase of the spin $S$ model where the bulk is pictorially given by the alternating $N_B^k=2S-(k-1)$ VB on $J_2$-link and $N_B^k+(k-1)$ VB on $J_1$-link. 
The open boundary condition corresponds to cutting $N_B^k$ VB ($J_2$-link), 
which induces effective $S^k_{\rm {\rm eff}}=N^k_B/2$ spins at both ends. 
Then the gap closing by $\Delta E_S(M_S,T/2)=0$ are for
$4S^k_{{\rm eff}}=2N^k_B=4S-2(k-1)$ different $M_S$ sectors \cite{ex1}.
It implies that the discontinuity at $t=T/2$ is
$-\sum_iP(t)\big|^{T/2+0}_{T/2-0}=-2S^k_{{\rm eff}}=-N^k_B$
(the sigh is determined by the way of exchange
of the left and right edge states). 
It contributes to the entanglement entropy by
$\log (2S_{{\rm eff}}+1)$ \cite{Katsura2007}.
Since this effective spin is spherical 
due to the symmetrization according to the VBS picture.

This scenario is consistently confirmed by numerical calculations. 
See Fig.~\ref{Fig2} (a), (b) and (c). 
For $S=1,3/2,2$ at $t=T/2$, the degeneracy specified by $\Delta E_S$ are two, four, and six-fold, respectively.
Further, for $S=2$ system at $t=T/2$, the level structure
due to spherical nature of the effective spin is discussed in detail (See Sec.IV in \cite{Sup}).

\begin{figure}[t]
\centering
\includegraphics[width=8cm]{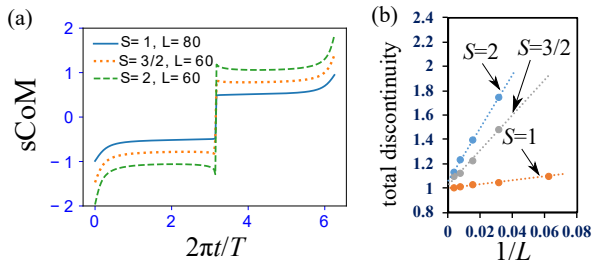}
\caption{
(a) Behavior of the sCoM with $S^{z}=0$ sector. For $S=1$ case, we set $L=80$, for $S=3/2$ and $2$, $L=60$. (b) System size dependence of the total discontinuity of the sCoM. It converges to an integer 
$I_{12}=\lim_{L\to \infty}\biggl[-\sum_iP(t)\big|^{t_i+\delta t}_{t_i-\delta t}\biggr]$.
In the numerical calculation of the discontinuity at $t=T/2$ in $S=1$, $3/2$ and $2$ cases, we set $\delta t=0.5\times 10^{-2}T$, $0.5\times 10^{-3}T$, and $0.5\times 10^{-3}T$.}
\label{Fig3}
\end{figure}
Then the total discontinuity of the pump protocol starting from SPT$k$ and passing through SPT$k ^\prime $ is given by $I_{k k ^\prime } = N^{k}_B-N^{k ^\prime }_B=k ^\prime -k$.
This is also consistently confirmed by numerical calculations. 
See Fig.~\ref{Fig3} (a), the behaviors of sCoM of the pump for the same parameters \cite{remark2} in Fig.~\ref{Fig2} (a), (b) and (c). For $S=1, 3/2, 2$ cases, the discontinuities of $P(t=T/2)$ numerically obtained are very close to $1,2,3$ $(=N^{k=2}_B)$. 
The total discontinuity of the pump protocol starting from SPT1 ($k=1$) and passing through SPT2 ($k'=2$) are given by $1(=k'-k)$. The numerical exploration to the infinite size is shown in Fig.~\ref{Fig3} (b) \cite{remark7}. 
It implies the total discontinuity of the sCoM approaches to 1 for $L\to \infty$, that is, $I_{12}=1$. 

\section{Bulk-edge correspondence, VBS and Berry phases}
The sCoM, $P(t)$, is a piecewise continuous function, that is, continuous
except several discontinuities at $t=t_i$ due to the appearance of effective boundary spins. 
This pumped spin by the continuous part,
is given by the bulk and is given by the Chern number as \cite{Hatsugai2016} (See also Sec.I in \cite{Sup}), 
$Q^e_{kk ^\prime } = \frac {1}{2\pi} \int_0^{2\pi} d \theta \, \bar Q_{kk ^\prime}^b(\theta )=C_{kk ^\prime}$, 
where  $\bar Q^b_{kk ^\prime }=
i\int_0^Tdt\, \partial_t \bar A_\theta^{(t)} (\theta)$ 
and $\bar A_\theta ^{(t)}$ is the 
Berry connection of the periodic system
in the temporal gauge $\bar A_t ^{(t)}=0$ where the twist $\theta$ is
distributed uniformly for all links.
This uniform twist is transformed to the local boundary twist in the Chern number 
 by the large gauge transformation, which makes the Chern number invariant \cite{Hatsugai2016,Sup}. 

Eqs.~(\ref{eq:bulk-charge}) and the representation of $Q^{e}_{kk'}$ imply
the bulk-edge correspondence for the generic quantum spins as
\begin{eqnarray*}
C_{kk ^\prime } &= I_{kk ^\prime }= k ^\prime -k.
\label{BEC}
\end{eqnarray*}
It also imposes a constraint for the Berry phases $\gamma^{k}-\gamma^{k'} \equiv \pi C_{kk ^\prime },\ \text{mod}\:2\pi$.
Here we have established the bulk-edge correspondence and discussed the numerical results based on the VBS picture. 
Reversely the topological stability of the Chern number
implies the (fractionalized) effective $S_{\rm eff}=S/2$ free spins at both ends are topologically stable. These effective edge spins are emergent and is topologically protected by the bulk gap. 
It results in inevitable level crossing of the pump with the open boundary condition as an emergent super-selection rule of the infinite system.

In \cite{Hatsugai2016} and in Sec.~I in \cite{Sup}, details of the topological pump and the bulk-edge correspondence are given. The large gauge transformation associated with the local $U(1)$ gauge transformation is fundamentally important.

\section{Conclusion}
We have clarified the presence of plateau transitions and a nontrivial topological spin pump in dimerized Heisenberg models ($S=1,3/2,2,3$) with the symmetry breaking term as a synthetic dimension. The model can be feasible for a recent coldatom system \cite{exp_realize}. 
The critical points between the various SPT phases are topological obstructions for the spin pump.
These obstructions protect the quantization of the Chern number
as the total pumped spin. 
Due to the bulk-edge correspondence which we have demonstrated, 
quantization of the Chern number implies emergent boundary degrees of freedom for the spin chains, 
which is consistent with the VBS picture of the dimerized Heisenberg model. 

Some high-dimensional systems related to our spin pump, such as a 2D topological charge pump \cite{Price2015}, simulated in a recent experiment \cite{Lohse2018}. The high-dimensional extension of our spin pump is a future interesting topic. Also, an extension to the $SU(N)$ spin chains (See \cite{PRL-TK-TM-YH} and \cite{Affleck1985}) can be straightforward.

\section*{Acknowledgments.---}
The authors thank T. Yoshida
for valuable discussions. The work is supported by JSPS KAKEN-HI Grant Number JP17H06138.


\clearpage
\renewcommand{\theequation}{S\arabic{equation}}
\renewcommand{\thefigure}{S\arabic{figure}}
\renewcommand{\bibnumfmt}[1]{[S#1]}
\renewcommand{\citenumfont}[1]{S#1}
\setcounter{equation}{0}
\setcounter{figure}{0}
\widetext
\section*{Supplemental Material: Plateau Transitions of Spin Pump and Bulk-Edge Correspondence}

\subsection*{I. Meaning of Chern number and Bulk edge correspondence of the spin pump}
In this section, we show that the Chern number is given by a bulk pumped spin and the Chern number in the periodic system with twist is related to the discontinuity of the sCoM, obtained in the open boundary system without twist.

Let us start a modified Hamiltonian of the system with open boundary as 
\begin{align*}
  H^{e}(\theta ) &= \sum_{j=1}^{L-1}\frac {1}{2} (
  e^{ -i\theta/L}S_{j}^+S_{j-1}^- +   e^{ +i\theta/L}S_{j}^-S_{j-1}^+)
  +\cdots .
\end{align*}
The current operator is given
\begin{align*}
  {\cal J} ^{e} &=\partial _ \theta H^{e} = - \frac {i}{L}
  \sum_{j=1}^{L-1}\frac {1}{2} (
  e^{- i\theta/L}S_{j}^+S_{j-1}^- -   e^{ i\theta/L}S_{j}^-S_{j-1}^+). 
\end{align*}
Then by the adiabatic approximation, the expectation value of the current
is
\begin{align*}
  J ^{e}&= \langle {\cal J} ^{e} \rangle _t = -i B^{e},
\end{align*}
where Berry curvature is $B^{e} = \partial _\theta A_t-\partial _t A_\theta$, 
and the Berry connection is $A_\alpha = \langle g|\partial _\alpha g \rangle$, and $|g\rangle$ is a snap shot ground states defined by $H^{e} | g \rangle = | g \rangle E$.
Especially in the temporal gauge $A_t^{(t)}=0$, it is
\begin{align*}
  J ^{e}&= +i \partial _t A^{(t)}_\theta ,
\end{align*}
where $A_\theta ^{(t)}$  is given by the Berry connection $A_\alpha $
in arbitrary gauge as
\begin{align*}
  A_\theta ^{(t)}(\theta ,t ) &= A_\theta (\theta ,t)-A_\theta (\theta ,0)-
  \partial _\theta \int _0^t d\tau\, A_t(\theta ,\tau).
\end{align*}
Let us introduce the large gauge transformation, 
\begin{align*}
U (\theta) &= e^{-i \theta \sum_{j=0}^{L-1}\frac {j-j_0}{L} S_j^z}, \:\: j_0 =\frac {L-1}{2}.
\end{align*}
The Hamiltonian of the open system is transformed as
\begin{align*}
  H^{e}(\theta )&=U H^{e}_0U ^{-1}, \  H^{e}_0=H^{e}(\theta =0),
\end{align*}
where $H_0^e$ is $\theta$ independent.

Now the snap shot ground states satisfy following relations
\begin{align*}
  H^{e}(\theta )|g \rangle &= E | g \rangle,
  \\
  H^{e}_0|g_0 \rangle &= E | g_0 \rangle,
  \\
  |g \rangle &= U| g_0 \rangle .
\end{align*}
Since  $|g_0 \rangle  $ is $\theta $ independent, using this gauge,
we have the following Berry connections
\begin{align*}
  A_t  &= \langle g| \partial _t  g \rangle =
  \langle g_0| \partial _t  g_0 \rangle :\theta\text{-independent},
  \\
  A_\theta  &= \langle g| \partial _\theta  g \rangle =
  \langle g_0|U ^{-1}   \partial _t  U | g_0 \rangle =-i P^e(t),
\end{align*}
where $P^e(t)$ is the sCoM for open system without twist $\theta$, 
\begin{align*}
  P^{e}(t) &= \langle g_0|
\sum_{j=0}^{L-1}\frac {j-j_0}{L} S_j^z  |g_0 \rangle.
\end{align*}
This sCoM $P^{e}(t)$ is calculated in the main text. 

The above relation implies the Berry connection in the temporal gauge $A_\theta ^{(t)}$
is given by
\begin{align*}
A_\theta ^{(t)}(\theta ,t ) &= A_\theta (\theta ,t)-A_\theta (\theta ,0)
=-i\big( P(t)-P(0)\big).
\end{align*}
Then the current for the open system is given by
\begin{align*}
J^{e} &= i \partial _t A_\theta ^{(t)}=\partial _t P  ^{e}.
\end{align*}
Now the pumped spin $Q^e_{[t_a,t_b]}$ (of the system with edges) for the period $[t_a,t_b]$ is give by
\begin{align*} 
Q^e_{[t_a,t_b]} &= \int_{t_a}^{t_b}dt\, J^{e} = P^{e}(t_b)-P^{e}(t_a).
\end{align*}
However due to the periodicity in time $T$, total pumped spin is zero and is
written as 
\begin{align*}
  0 &= \int_{0}^T dt\, \partial_t P^{e}
  =\sum_i\int_{t_{i-1}}^{t_i}dt\, \partial_t P^{e} +P^{e}(t)\bigg|^{t_i+0}_{t_i-0}
\end{align*}
where $P^e$ is singular and discontinuous at $t=t_i$, ($i=1,2,\cdots$)
(periodicity in time is assumed for the summation). 

Since the contribution to the pumped spin with edges
from continuous parts are due to bulk, 
the pumped charge due to bulk is defined as 
\begin{align*}
   Q^e &\equiv   \sum_i\int_{t_{i-1}}^{t_i}dt\, \partial_t P^{e}.
\end{align*}
The subscript $^e$ implies that  it is for a system with edges. 
Note that it is determined only by the information of the system with edges as
\begin{align*}
  Q^e &= I,
\end{align*}
where
\begin{align*}
  I =-\sum_i  P^{e}(t)\bigg|^{t_i+0}_{t_i-0},
\end{align*}
is a sum of the discontinuity due to edges. In the main text, we numerically estimated the sum of discontinuity in Fig.~3 (b).

Physically this pumped charge (spin) of bulk is related to the current of a periodic system,
${\cal J}^b=i\partial _t \bar A_\theta ^{(t)}$. 
{\it We can safely assume} the pumped spin of the bulk, $Q^b$ as  
\begin{align*}
   Q^e &= 
 \frac {1}{2\pi} \int_0^{2\pi} d \theta \, 
 \bar Q^b (\theta ),
\end{align*}
where 
\begin{align*}
\bar Q^b (\theta )&=
i\int_0^Tdt\, \partial _t \bar A_\theta ^{(t)}.
\end{align*}
$\bar Q^b (\theta )$ is given by a temporal gauge Berry connection $\bar A_\theta ^{(t)}$.  

The temporal gauge Berry connection $\bar A_\theta ^{(t)}$ is obtained by the Berry connection $\bar A_\alpha $ (arbitrary gauge) of the periodic system described by the Hamiltonian ($S_0^{x,y,z} =S_L^{x,y,z}$)
\begin{align*}
  \bar H^{b}(\theta ) &= \sum_{j=1}^{L}\frac {1}{2} (
  e^{ -i\theta/L}S_{j}^+S_{j-1}^- +   e^{ +i\theta/L}S_{j}^-S_{j-1}^+)
  +\cdots
  \\
  &=
  H^{e}(\theta ) +
  \frac {1}{2} (
  e^{ -i\theta/L}S_{0}^+S_{L-1}^- +   e^{ +i\theta/L}S_{0}^-S_{L-1}^+), 
  \\
  \bar A_\alpha &= \langle \bar g^{b}| \partial _\alpha \bar g^{b} \rangle,
\end{align*}
where $|g^{b}\rangle$ is a snap shot ground states, $\bar H^{b}| \bar g^{b} \rangle = | \bar g^{b} \rangle E$, 
and also we have introduced an average over $\theta $, which is justified
if the system size is sufficiently large \cite{FHS2005,Kudo}.

From the above relation of $Q^{e}$ and ${\bar Q}^{b}(\theta)$, it establishes the bulk-edge correspondence
\begin{align*}
  I &= C.
\end{align*}
Here, the Chern number $C$ is for the periodic system as
\begin{align*}
C &=  \frac {i}{2\pi} \int_0^Tdt\,\int_0^{2\pi}d \theta \, \partial _t
\bar A^{(t)}_\theta =
\frac {1}{2\pi i} \int_0^Tdt\,\int_0^{2\pi}d \theta \, \bar B,
\\
\bar B &= \partial _\theta \bar A_t -\partial _t \bar A_\theta,
\end{align*}
where $\bar A_\alpha $ is in arbitrary gauge.

By considering $\bar H^{b}(\theta)$, which has uniform twist $\theta/L$, we showed the bulk-edge correspondence $I=C$.
However, a periodic boundary system, which has one link twist $\theta$, used in the main text, also gives Chern number equivalent to that of $\bar H^{b}(\theta)$. From now on, we show it.

Similar to the open system, one may perform the same 
unitary transformation $U$, and define $H^b_0$
as
\begin{align*}
  \bar H^{b} &=  U H^{b}_0 U ^{-1}.
\end{align*} 
And we define a snap shot ground state, $H^{b}_0 |g_0^b \rangle = |g_0^b \rangle E$, which is related to $|g^{b}\rangle$ as $|\bar g^b \rangle = U|g_0^b \rangle$. 
Although $H^e_0$ is $\theta $ independent, $H^b_0$ depends on
$\theta $ as 
\begin{align*}
  H^b_0 &= H_0^e+
  \frac {1}{2} (
  e^{i\theta}S_{0}^+S_{L-1}^- + e^{-i\theta}S_{0}^-S_{L-1}^+) . 
\end{align*}
The last term is the boundary twist which we have used for the calculation of
the Chern number/ Berry phase in the main text.
Then, the Berry connections, $\bar A_\alpha $ and
$\hat A_\alpha $ for the $\bar H^b$ and $H^b_0$ are related by
\begin{align*}
  \bar A _ \theta &= 
  \langle \bar g^{b}| \partial _\theta \bar g_0^{b} \rangle
  =
  \langle  g_0^{b}| \partial _\theta  g_0^{b} \rangle  +
  \langle  g_0^{b}|U ^{-1}  \partial _\theta U| g_0^{b} \rangle
  = \hat A_\theta -i  \hat P^b(t),
  \\
  \bar A_t &=   \hat A_t,
\end{align*}
where
$\hat A_\alpha  =   \langle  g_0^{b}| \partial _\alpha g^{b}_0 \rangle$
and $\hat P(t,\theta ) = \langle g_0^b|\sum_{j=0}^{L-1}\frac {j-j_0}{L} S_j^z| g_0^b \rangle$.
$\hat P(t,\theta )$ is an effective sCoM for the bulk, that depends on
the choice of the origin although the system is
periodic. 
{\it Therefore it is unphysical and does not have any experimental significance.} Note that $\hat P(t,\theta )$ is a smooth function of $t$ and $\theta $ with periods $T$ and $2\pi$, although $P(t)$ for the system with boundaries is singular.
Since $\hat P$ is smooth and periodic, the Chern numbers
by the Berry connections $\hat A_\alpha $ and $\bar A_\alpha $
are the same (their field strength are different).
It is explicitly written as
\begin{align*}
  \hat B &= \partial _\theta \hat A_t - \partial _t  \hat A_\theta,
 \\
  \bar B &= \partial _\theta \bar A_t - \partial _t  \bar A_\theta  
  =\hat B  -i\partial _t \hat P^b,
  \\
  C &= \frac {1}{2\pi i} \int_{0}^{T}dt\,\int_{0}^{2\pi} d\theta\, \bar B
  = \frac {1}{2\pi i} \int_{0}^{T}dt\,\int_{0}^{2\pi}d \theta\, \hat B.
\end{align*} 
Hence, Chern number obtained from $H^{b}_0$ is equivalent to that of $\bar H^{b}(\theta)$.

\subsection*{II. $S=3$ plateau transition}

\begin{figure}[h]
\centering
\includegraphics[width=7cm]{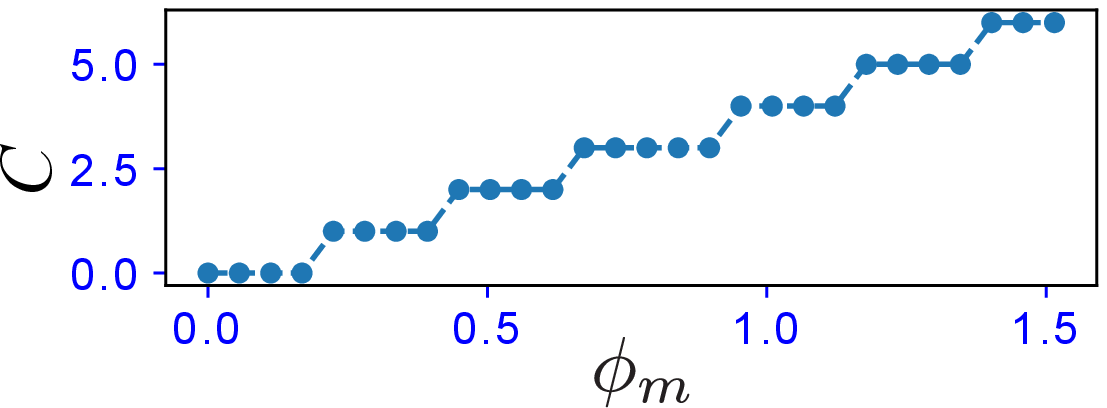}
\caption{Plateau transitions for $S=3$ case.}
\label{FigS1}
\end{figure}

From the dimerized Heisenberg model of $H_{DH}$ of Eq.~(1) in the main text, we expect that $2S+1$ SPT phases appear for arbitrary $S$ case. By introducing the symmetry breaking term, $S(2S+1)$ pump loops with the nontrivial topological pump are allowed and the highest Chern number is $C=2S$.
For $S=3$ case, we consider the dimerized Heisenberg model of $H_{DH}$ of Eq.~(1) in the main text with $J_1=\sin\phi$, $J_2=\cos\phi$. 
Here, we expect that in $\phi \in [0,\pi/2]$, the model exhibits $2S+1=7$ SPT phases. Then, we set an pump protocol, 
$J_1=\sin\phi$, $J_2=\cos\phi$, $\phi=\phi_{m}[1-\cos(2\pi t/T)]/2$ 
and $\Delta(t)=\sin(2\pi t/T)$.  
For this protocol, we calculated the Chern number $C$ of Eq.~(2) in the main text by using the exact diagonalization \cite{Quspin}. 
The $\phi_m$ dependent result of $C$ is shown in Fig.~\ref{FigS1}, where we set system size $L=10$. 
There is also no system size dependence.  
We observe plateau transitions again. 
The highest Chern number is $2S=2\times 3=6$. 
Accordingly, we expect for further higher integer $S$ case, further plateau transition behavior with more steps emerges.

\subsection*{III. Quantization and plateau transition of Chern number from SPT transition line picture}

In the main text, we showed how to quantize the Chern number. 
As an additional picture of the quantization mechanism, in this supplementary, we also argue the quantization mechanism by considering the presence interaction.

\begin{figure}[h]
\centering
\includegraphics[width=18cm]{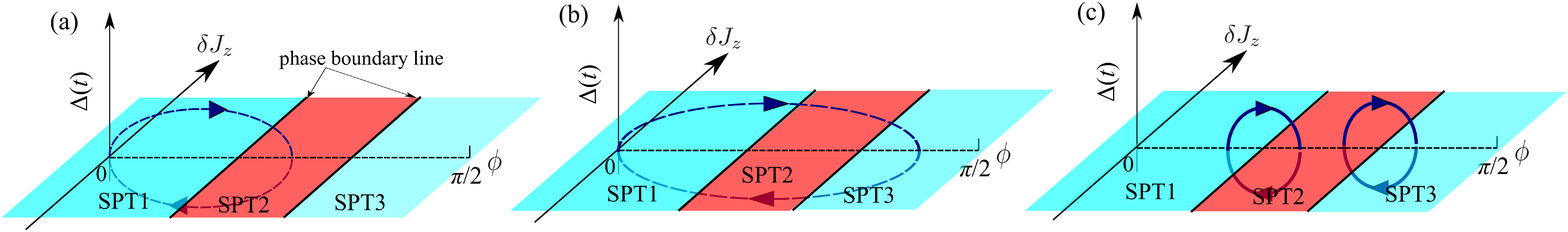}
\caption{Schematic figures for the mechanism of the quatization of the topological pump; 
(a) A pump protocol connects the SPT1 and SPT2 phases and wraps the single phase boundary line. 
(b) A pump protocol connects the SPT1 and SPT3 phases and wraps two phase boundary lines. (c) Pump protocol loop of (b) case can be adiabatically deformed into two loops.}
\label{FigS2}
\end{figure}

Let us introduce a perturbation, $H_{zz}=\delta J_z\sum^{L-1}_{j=0}S^{z}_{j+1}S^{z}_j$, where $\delta J_{z}$ is small. 
The term $H_{zz}$ does not break the symmetries of the SPTs. 
Accordingly, when one considers $\Delta-\phi-\delta J_z$ space for the system of $H_{DH}+H_{SB}+H_{zz}$ with small $\delta J_{z}$, the critical transition points on $\phi$ space form phase boundary lines separating the SPTs on $\phi-\delta J_z$ plane 
($
\Delta=0
$) as shown in Fig.~\ref{FigS2}. For a finite small $\delta J_z$, each SPT phases are stable and remain gapped. For an infinite system, the phase boundary lines are gapless for any twist $\theta$. 

For this model parameter space, one sets the pump protocol defined previously. 
As shown in Fig.~\ref{FigS2} (a), if one considers the pump protocol with $\phi_m=\pi/4$ for $S=1$ system defined in the main text, the pump protocol wraps the single phase boundary line separating the SPT1 and SPT2. One cannot untie the pump protocol loop and the phase boundary line without gap closing. That is, the phase boundary line can be regarded as a topological obstruction. 
Then the calculation of $C$ on the pump protocol gives $C=1$. 
In other words, the Chern number may be given by a linking number of a pump protocol and the phase boundary lines. 
One can also set a larger pump protocol wrapping two phase boundary lines as shown in Fig.~\ref{FigS2} (b). 
Then, one obtains $C=2$.
This pump can be also understood by deforming the loop of the pump protocol.
The loop can be adiabatically deformed to two loops wrapping a single phase boundary line without gap closing as shown in Fig.~\ref{FigS2} (c). 
Each loop gives $C=1$. Thus, one can consider a sum rule of $C$, obtain the total Chern number, $C=2$. 
The number of the phase boundary line in a pump protocol corresponds to the quantized value of $C$. From these arguments, increasing $\phi_m$ corresponds to the increase of the loop size of the pump protocol. Then the number of the phase boundary line on $\phi-\delta J_z$ plane wrapped by the pump protocol increases. When the number of the phase boundary line changes, a transition of Chern number occurs.

\subsection*{IV. Hybridization of edge states and $S=3/2$ edge state picture}
In Fig.2 (a) and (b) in the main text, 
we showed the excitation spectrum obtained by the ground state energies with different $S^{z}$. 
There, multiple ingap states appear. 
The ingap states are left or right edge states. 
In particular, at high symmetry points $t=0$ and $T/2$, where $H_{SB}$ in the main text vanishes, some ingap states are degenerate.  
Here, we further study the spectrum structure from the presence of edge states. 
We focus on the spectrum of $S=2$ case at $t=T/2$ as shown in Fig.2 (b) in the main text. 
For a finite-size system with open boundary,
at $t=T/2$, the valence-bond-solid (VBS) picture predicts the presence of $S=3/2$ spin at both edges. 
We investigate whether or not such a $S=3/2$ spin picture at both edges is correct. 


In an old study \cite{Kennedy}, $S = 1$ Heisenberg model with a biquadratic term ($S=1$ bilinear-biquadratic spin chain) are studied to show adiabatic connection between the the AKLT model and the simple Heisenberg model. It justified the VBS picture of the bulk and existence of the boundary  $S = 1/2$ degrees of freedom that contribute to the low energy spectrum in the Haldane gap.
For the finite-size system with open boundary,
the nearly degeneracy of the spectrum coming from the hybridization of edge $1/2$-spins was clarified, where the lowest excited states are called ``Kennedy triplet'' and these states becomes degenerate to the groundstate energy for infinite system size \cite{Kennedy}.

\begin{figure}[h]
\centering
\includegraphics[width=13cm]{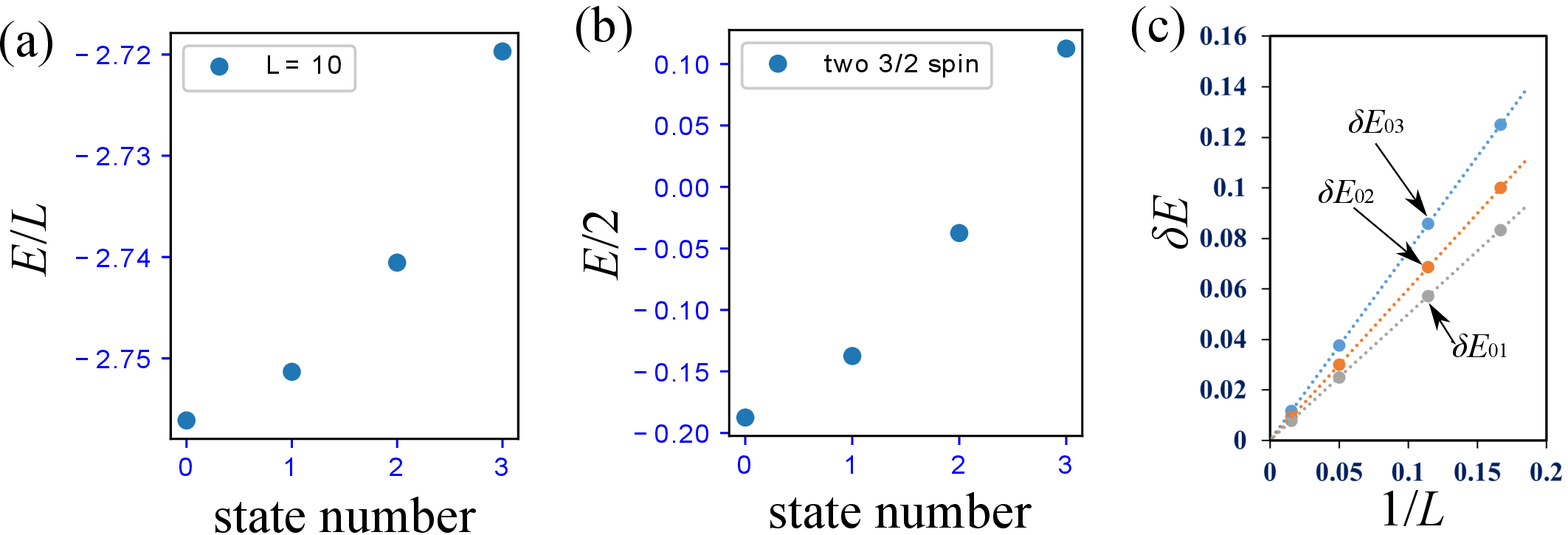}
\caption{(a) Low-lying spectrum structure of $S=2$ exact diagonalization. (b) The spectrum structure of the two $3/2$-spin system, $H_{e}$ with zero magnetization sector. (c) System size dependence of energy differences $\delta E_{jk}$.}
\label{FigS3}
\end{figure}

Motivated by the old work \cite{Kennedy}, we calculated the low-lying energies for $S=2$ system in the main text with open boundary at $t=T/2$ with $S^{z}=0$ sector by using \cite{Quspin}. 
The four lowest energies are plotted in Fig.~\ref{FigS3} (a). 
The four energies are split due to finite system size effects. 
If the edge state of the system is described by $S=3/2$ spin, then we expect that the spins at both edges interact with each other with a small effective coupling $J_e$. 
We expect that, phenomenologically, $J_e$ is $\propto e^{-L/\xi}$, where $L$ is a system size and $\xi$ is a correlation length.
The coupling $J_e$ induces hybridization of the edge states. 
The hybridization is determined by $S=3/2$ two-spin system. 
The two-spin system is given by an effective edge spin Hamiltonian, $H_{e}=J_{e}\vec{s}_{L}\cdot \vec{s}_{R}$, where $\vec{s}_{L(R)}$ is a left (right) $S=3/2$ spin. Here, by setting $J_e=0.1$, we plot the spectrum $H_{e}$ with zero-magnetization sector. The result is shown in Fig.~\ref{FigS3} (b).

Let us compare the spectrum structures of Fig.~\ref{FigS3} (a) with that of Fig.~\ref{FigS3} (b). Both structures are much similar. 
To estimate the similarity quantitatively, we consider a level spacing ratio, $r_{1}=\frac{|E_0-E_1|}{|E_1-E_2|}$ and $r_{2}=\frac{|E_1-E_2|}{|E_2-E_3|}$, where $E_{i}$ ($i=0,1,2,3$) is $i$-th energy in ascending order. 
We calculated $r_1$ and $r_2$ for $S=2$ system with open boundary at $t=T/2$ with $S^{z}=0$ sector. We set different system size $L=6,8,10$. 
Then, the results for $L=6,8,10$ are $(r_1,r_2)\equiv (r^{ED}_{1},r^{ED}_2)=(0.44977,0.53621), (0.44648,0.52350), (0.44497,0.51708)$. 
The system size dependence is small. 
On the other hand, for the $S=3/2$ two-spin system of $H_{e}$ with zero-magnetization sector, $(r_1,r_2)\equiv (r^e_{1},r^e_{2})=(1/2,2/3)$. 
Therefore, $r^{ED}_{1}$ and $r^{ED}_2$ are close to $r^{e}_1$ and $r^e_2$. 
From this comparison, the low-lying spectrum of $S=2$ spin chain at $t=T/2$ can be effectively described by $S=3/2$ spin picture at both edges. That is, the VBS picture as shown in Fig.1 (c) is valid. 

Finally, we show that the low-lying spectrum for the $S=2$ system comes to be degenerate. Figure \ref{FigS3} (c) shows the system size dependence of energy differences defined by $\delta E_{jk}=|E_{j}-E_{k}|/L$.
The differences approach to zero for $L\to \infty$. 
This result implies that the groundstate for the $S=2$ system becomes four-fold degenerate for $L\to \infty$. 
This result is analogous to the result for $S=1$ Haldane phase in \cite{Kennedy}.

\begin{thebibliography}{99}
\bibitem{Haldane}
F. D. M. Haldane, Rev. Mod. Phys. {\bf 89}, 040502 (2017).

\bibitem{Kennedy}
T. Kennedy, J. Phys.: Condens. Matter {\bf 2}, 5737 (1990).

\bibitem{Hatsugai1992}
Y. Hatsugai, J. Phys. Soc. Jpn. {\bf 61}, 3856 (1992).

\bibitem{White}
S. R. White and D. A. Huse, Phys. Rev. B {\bf 48}, 3844 (1993).

\bibitem{Nakano2009}
H. Nakano and A. Terai, J. Phys. Soc. Jpn. {\bf 78}, 014003 (2009).

\bibitem{Nakano2018}
H. Nakano, and T. Sakai, J. Phys. Soc. Jpn. {\bf 87}, 105002 (2018).

\bibitem{Affleck1987}
I. Affleck, T. Kennedy, E. H. Lieb, and H. Tasaki, Phys. Rev. Lett. {\bf 59}, 799 (1987).

\bibitem{Affleck1988}
I. Affleck, T. Kennedy, E. Lieb, and H. Tasaki, Commun. Math. Phys. {\bf 115}, 477 (1988).

\bibitem{Affleck1989}
I. Affleck, J. Phys.: Condens. Matter {\bf 1}, 3047 (1989).

\bibitem{Hagiwara}
M. Hagiwara, K. Katsumata, I. Affleck, B. I. Halperin, and J. P. Renard, Phys. Rev. Lett. {\bf 65}, 3181 (1990).

\bibitem{Hatsugai1993}
Y. Hatsugai, Phys. Rev. Lett. {\bf 71}, 3697 (1993).

\bibitem{Pollmann2010}
F. Pollmann, A. M. Turner, E. Berg, and M. Oshikawa, 
Phys. Rev. B {\bf 81}, 064439 (2010).

\bibitem{Chen}
X. Chen, Z.-C. Gu, and X.-G. Wen, Phys. Rev. B {\bf 84}, 235128 (2011).

\bibitem{Pollmann2012}
F. Pollmann, E. Berg, A. M. Turner, and M. Oshikawa, Phys. Rev. B {\bf 85}, 075125 (2012).

\bibitem{Hirano2008_2}
T. Hirano, H. Katsura, and Y. Hatsugai, Phys. Rev. B {\bf 78}, 054431 (2008).

\bibitem{Hirano2008}
T. Hirano, H. Katsura, and Y. Hatsugai, Phys. Rev. B {\bf 77}, 094431 (2008). 

\bibitem{Katsura2007}
H. Katsura, T. Hirano, and Y. Hatsugai, Phys. Rev. B {\bf 76}, 012401 (2007).

\bibitem{Mila}
N. Chepiga, F. Michaud, and F. Mila, Phys. Rev. B {\bf 88} 184418 (2013).

\bibitem{Fubasami}
S. Fubasami, T. Mizoguchi, and Y. Hatsugai, Phys. Rev. B {\bf 100}, 014438 (2019).

\bibitem{Thouless}
D. J. Thouless, Phys. Rev. B {\bf 27}, 6083 (1983).

\bibitem{Lohse} 
M. Lohse, C. Schweizer, O. Zilberberg, M. Aidelsburger, and I. Bloch, Nat. Phys. {\bf 12}, 350 (2016).

\bibitem{Nakajima}
S. Nakajima, T. Tomita, S. Taie, T. Ichinose, H. Ozawa, L. Wang, M. Troyer, 
and Y. Takahashi, Nat. Phys. {\bf 12}, 296 (2016).

\bibitem{Schweizer}
C. Schweizer, M. Lohse, R. Citro, and I. Bloch, 
Phys. Rev. Lett. {\bf 117}, 170405 (2016).

\bibitem{Kraus_ex}
Y. E. Kraus, Y. Lahini, Z. Ringel, M. Verbin, and O. Zilberberg, Phys. Rev. Lett. {\bf 109}, 106402 (2012).

\bibitem{Ozawa}
T. Ozawa, H. M. Price, A. Amo, N. Goldman, M. Hafezi, L. Lu, M. Rechtsman, D. Schuster, J. Simon, O. Zilberberg, and I. Carusotto, 
Rev. Mod. Phys. {\bf 91}, 015006 (2019).

\bibitem{Cooper}
N. R. Cooper, J. Dalibard, and I. B. Spielman, Rev. Mod. Phys. {\bf 91}, 015005 (2019).

\bibitem{Hatsugai2016}
Y. Hatsugai and T. Fukui, Phys. Rev. B {\bf 94}, 041102(R) (2016).

\bibitem{KH2020}
Y. Kuno and Y. Hatsugai, Phys. Rev. Res. {\bf 2} 042024(R) (2020). 

\bibitem{Po-Shen2020}
{P. S. Hsin, A. Kapustin, and R. Thorngren, Phys. Rev. B {\bf 102}, 245113 (2020).}

\bibitem{Wang}
L. Wang, M. Troyer, and X. Dai, Phys. Rev. Lett. {\bf 111}, 026802 (2013).

\bibitem{RLi}
R. Li and M. Fleischhauer, Phys. Rev. B {\bf 96}, 085444 (2017).

\bibitem{YKe}
Y. Ke, X. Qin, Y. S. Kivshar, and C. Lee, Phys. Rev. A {\bf 95}, 063630 (2017).

\bibitem{Kuno2017}
Y. Kuno, K. Shimizu, and I. Ichinose, New J. Phys. {\bf 19}, 123025 (2017).

\bibitem{Nakagawa}
M. Nakagawa, T. Yoshida, R. Peters, and N. Kawakami, Phys. Rev. B {\bf 98}, 115147 (2018).

\bibitem{Hayward}
A. Hayward, C. Schweizer, M. Lohse, M. Aidelsburger, and
F. Heidrich-Meisner, Phys. Rev. B {\bf 98}, 245148 (2018).

\bibitem{Greschner}
S. Greschner, S. Mondal, and T. Mishra, Phys. Rev. A {\bf 101}, 053630 (2020).

\bibitem{Shindou}
R. Shindou, J. Phys. Soc. Jpn. {\bf 74}, 1214 (2005).

\bibitem{Hu}
H. Hu, C. Cheng, Z. Xu, H. G. Luo, and S. Chen, Phys. Rev. B {\bf 90}, 035150 (2014).

\bibitem{TKNN}
D. J. Thouless, M. Kohmoto, P. Nightingale, and M. den Nijs, Phys. Rev. Lett. {\bf 49}, 405 (1982).

\bibitem{Kraus}
Y. E. Kraus, Y. Lahini, Z. Ringel, M. Verbin, and O. Zilberberg, Phys. Rev. Lett. {\bf 109}, 106402 (2012).

\bibitem{Nakamura}
M. Nakamura and S. Todo, Phys. Rev. Lett. {\bf 89}, 077204 (2002).

\bibitem{exp_model}
The model is related to the bilinear-biquadratic $S=1$ chains, the experimental construction method of which has been proposed: A. Imambekov, M.Lukin, E. Demler, Phys. Rev. A {\bf 68}, 063602 (2003); J. J. Garcia-Ripoll, M. A. Martin-Delgado, and J. I. Cirac, Phys. Rev. Lett. {\bf 93}, 250405 (2004); G. K. Brennen, A. Micheli, and P. Zoller, New J. Phys. {\bf 9} 138 (2007).


\bibitem{Yajima}
M. Yajima, and M. Takahashi, J. Phys. Soc. Jpn. {\bf 65}, 39 (1996). 

\bibitem{Yamamoto}
S. Yamamoto, Phys. Rev. B {\bf 55}, 3603 (1997).

\bibitem{Kitazawa}
A. Kitazawa and K. Nomura, J. Phys. Soc. Jpn. {\bf 66} 3944 (1997).

\bibitem{remark_symmetry}
$D_{2}$ transformation means that ${\bf Z}_2\times {\bf Z}_2$ spin rotation, that is, spin $\pi$ rotation with respect to any two of spin $x, y, z$ axis. Time-reversal transformation means ${\vec S}_j\to -{\vec S}_j$. Bond-centered-inversion transformation means $\vec{S_j}\to\vec{S}_{L-j-1}$ ($L$ is a system size).

\bibitem{EPL-YHIM}
Y. Hatsugai and I. Maruyama, Europhys. Lett. {\bf 95}, 20003 (2011).

\bibitem{Hatsugai2005}
Y. Hatsugai, J. Phys. Soc. Jpn. {\bf 74}, 1374 (2005).

\bibitem{Hatsugai2006}
Y. Hatsugai, J. Phys. Soc. Jpn. {\bf 75}, 123601 (2006).

\bibitem{Hatsugai2007}
Y. Hatsugai, J. Phys. Condens. Matter {\bf 19}, 145209 (2007).

\bibitem{Hatsugai2011}
Y. Hatsugai and I. Maruyama, Europhys. Lett. {\bf 95}, 20003 (2011).

\bibitem{PRL-TK-TM-YH}
T. Kariyado, T. Morimoto, and Y. Hatsugai, Phys. Rev. Lett. {\bf 120}, 247202 (2018).

\bibitem{Kudo}
K. Kudo, H. Watanabe, T. Kariyado, and Y. Hatsugai, Phys. Rev. Lett. {\bf 122}, 146601 (2019).

\bibitem{NTW}
Q. Niu, D. J. Thouless, and Y. -S. Wu, Phys. Rev. B {\bf 31}, 3372 (1985).

\bibitem{Arovas}
F. D. M. Haldane and D. P. Arovas, Phys. Rev. B {\bf 52}, 4223 (1995).

\bibitem{Sup}
See Supplemental Material; I. Meaning of Chern number and Bulk edge correspondence of the spin pump, II. $S=3$ plateau transition, 
III. Quantization and plateau transition of Chern number from SPT transition line picture, IV. Hybridization of edge states and $S=3/2$ edge state picture.

\bibitem{FHS2005}
T. Fukui, Y. Hatsugai, and H. Suzuki, J. Phys. Soc. Jpn. {\bf 74}, 1674 (2005).

\bibitem{Quspin}
We employed the Quspin solver: P. Weinberg and M. Bukov, SciPost Phys. {\bf 7}, 20 (2019); {\bf 2}, 003 (2017).

\bibitem{remark0}
All of the results are for system size $L=10$. 
There is no significant system size dependence.

\bibitem{Kivelson}
S. Kivelson, D. -H. Lee, and S. -C. Zhang, Phys. Rev. B {\bf 46}, 2223 (1992).

\bibitem{Hatsugai1999}
Y. Hatsugai, K. Ishibashi, and Y. Morita, Phys. Rev. Lett. {\bf 83}, 2246 (1999).

\bibitem{Kagalovsky}
V. Kagalovsky, B. Horovitz, Y. Avishai, and J. T. Chalker, Phys. Rev. Lett. {\bf 82}, 3516 (1999).

\bibitem{Kawarabayashi}
T. Kawarabayashi, Y. Hatsugai, and H. Aoki, Phys. Rev. Lett. {\bf 103}, 156804 (2009).

\bibitem{Morita}
Y. Morita and Y. Hatsugai, Phys. Rev. B {\bf 62}, 99 (2000).

\bibitem{remark10}
It is justified since one can deform the protocol loop to a sum of small loops around the critical points without gap closing. 
Here the Chern number of the critical point is given by the Chern number of the small pump around the point.

\bibitem{remark11}
When one extends the parameter space, by 1, respecting the symmetry of the SPT, these gapless points become phase boundary lines \cite{KH2020} (for the detail see Sec. III in \cite{Sup}).

\bibitem{TeNPy}
J. Hauschild and F. Pollmann, SciPost Phys. Lect. Notes {\bf 5} (2018).

\bibitem{remark5}
The left and right edge states are identified by the distribution of the local $z$-component magnetization 
$\langle g (t)|S^{z}_j| g (t)\rangle$.

\bibitem{ex1} 
For $S=1$ case in Fig.~\ref{Fig2} (a), at $t=0$ (SPT1 phase), 
$S_{eff}=1$ ($N^1_B=2$) spins appear at edges. For $4S^1_{eff}=4$ different $M_S$ sectors appears. The total degeneracy $(2S_{eff}+1)=(2+1)^2$-fold degenerate states.

\bibitem{remark2}
The numerical calculation of $P(t)$ is set in $S^{z}=0$ sector without the twist $\theta$.

\bibitem{remark7}
At $t=0$, the model is in the dimerized limit, $J_1=0$. 
Therefore, the discontinuity of the sCoM for $S=1$, $3/2$, and $2$ around $t=0$ with $S^z=0$ sector exactly becomes $-2$, $-3$ and $-4$, respectively.

\bibitem{exp_realize}
Much recently, a ladder optical lattice has realized the SPT phase \cite{Sompet2021}, that can be effectively $S=1$ Haldane chain. 
The observed SPT phase is related to the one of our spin model described in this letter.  
It implies that our target spin Hamiltonian $H$ and spin pump can be directly implemented and simulated by employing such a ladder optical lattice system with suitable fine-tuning.

\bibitem{Sompet2021}
P. Sompet, S. Hirthe, D. Bourgund, T. Chalopin, J. Bibo, J. Koepsell, P. Bojovic, R. Verresen, F. Pollmann, G. Salomon, C. Gross, T. A. Hilker, I. Bloch, 	arXiv:2103.10421 (2021).

\bibitem{Price2015}
H. M. Price, O. Zilberberg, T. Ozawa, I. Carusotto, N. Goldman, Phys. Rev. Lett. {\bf 115}, 195303 (2015).

\bibitem{Lohse2018}
M. Lohse, C. Schweizer, H. M. Price, O. Zilberberg, I. Bloch, Nature 553, 55 (2018).

\bibitem{Affleck1985}
I. Affleck, Phys. Rev. Lett. {\bf 54}, 966 (1985).


\end{thebibliography}

\begin{thebibliography}{99}
\bibitem{FHS2005}
T. Fukui, Y. Hatsugai, and H. Suzuki, J. Phys. Soc. Jpn. {\bf 74}, 1674 (2005).
\bibitem{Kudo}
K. Kudo, H. Watanabe, T. Kariyado, and Y. Hatsugai, Phys. Rev. Lett. {\bf 122}, 146601 (2019).
\bibitem{Quspin}
P. Weinberg and M. Bukov, SciPost Phys. {\bf 7}, 20 (2019); {\bf 2}, 003 (2017).
\bibitem{Kennedy}
T. Kennedy, J. Phys.: Condens. Matter {\bf 2} 5737 (1990).
\end{thebibliography}
\end{document}